\documentclass{article}
\usepackage[letterpaper,margin=1in]{geometry}
\usepackage{graphicx}
\usepackage[usenames,dvipsnames]{color}
\usepackage{authblk}
\usepackage{natbib}
    \setcitestyle{aysep={}} 

\usepackage[utf8]{inputenc}
\usepackage[T1]{fontenc}
\usepackage{amsmath,amssymb}
\usepackage[normalem]{ulem}

\usepackage{setspace}
\renewcommand{\d}{\mathrm d}
\renewcommand{\vec}[1]{\boldsymbol{#1}}
\newcommand{\R}{\mathbb R}
\newcommand{\norm}[1]{\left\| #1 \right\|}
\newcommand{\grad}{\boldsymbol\nabla}
\newcommand{\cross}{\times}
\newcommand{\pder}[2]{\frac{\partial #1}{\partial #2}}

\title{Direct, simple, and efficient computation of all components of the virtual-casing magnetic field in axisymmetric geometries with Kapur-Rokhlin quadrature}

\begin{document}

\author[a]{Evan Toler}
\author[b]{A. J. Cerfon}
\author[c]{D. Malhotra}

\affil[a]{Argonne National Laboratory, Lemont, IL 60439}
\affil[b]{Type One Energy Group - Canada Inc., Vancouver BC V6E 0A2, Canada}
\affil[c]{Flatiron Institute, New York, NY 10012}

\date{}

\maketitle

\begin{abstract}
In a recent publication \citep{toler_cerfon_malhotra_2023}, we demonstrated that for axisymmetric geometries, the Kapur-Rokhlin quadrature rule provided an efficient and high-order accurate method for computing the \textit{normal} component, on the plasma surface, of the magnetic field due to the toroidal current flowing in the plasma, via the virtual-casing principle. The calculation was indirect, as it required the prior computation of the magnetic vector potential from the virtual-casing principle, followed by the computation of its tangential derivative by Fourier differentiation, in order to obtain the normal component of the magnetic field. Our approach did not provide the other components of the virtual-casing magnetic field.  

In this letter, we show that a more direct and more general approach is available for the computation of the virtual-casing magnetic field. The Kapur-Rokhlin quadrature rule accurately calculates the principal value integrals in the expression for \textit{all} the components of the magnetic field on the plasma boundary, and the numerical error converges at a rate nearly as high as the indirect method we presented previously. 
\end{abstract}

\section{Introduction} \label{sec:intro}

The virtual-casing principle \citep{Shafranov_1972,Zakharov_1973,hanson2015virtual} is an effective tool for computing the magnetic field due to the currents flowing in the plasma, since it reduces the dimensionality of the integrals involved in this computation, from volume to surface integrals for non-axisymmetric geometries \citep{Lazerson_2013,malhotra2019efficient,kappel2023magnetic}, and from surface to line integrals for axisymmetric settings \citep{Zakharov_1973,ZakharovPletzer,toler_cerfon_malhotra_2023}. Computing the magnetic field with the virtual-casing principle is nevertheless challenging when the evaluation point is on the plasma boundary, because it involves the evaluation of integrals with singular integrands. 

Singularity subtraction schemes have often been used by the plasma physics community to address this difficulty \citep{Drevlak_2018}. They are robust, but are typically low-order accurate schemes \citep{malhotra2019efficient}, which are also tedious to implement. In contrast, there exist quadrature weights and abscissae in the applied mathematics literature which are specifically designed to give high accuracy for the singularities encountered in plasma physics applications, and are simple to implement. In recent work \citep{toler_cerfon_malhotra_2023}, we demonstrated that for axisymmetric geometries, the Kapur-Rokhlin quadrature scheme \citep{kapur1997high} provided high-order accuracy for the computation of the normal component on the plasma surface of the virtual-casing magnetic field. 
Since the Kapur-Rokhlin scheme is designed for integrable singularities with known asymptotic behavior, we constructed this calculation in two steps. We first computed the virtual-casing vector potential, which can be expressed in terms of an integral with a known logarithmic singularity. Then, we calculated the normal component of the magnetic field on the plasma surface by computing the tangential derivative of the vector potential, by Fourier differentiation. 

In the present letter, we show that the Kapur-Rokhlin quadrature scheme can be used in a more direct and more general way for the virtual-casing principle in axisymmetric geometries. Specifically, we demonstrate that the Kapur-Rokhlin scheme accurately computes the Cauchy principal value integrals arising in the expressions for the magnetic field in the virtual-casing principle, allowing us to compute all components of the magnetic field directly anywhere on the plasma boundary, via line integrals that are straightforward to implement and do not require any manipulation of the integrand. 

Our previous article and the present letter address different needs, associated with different applications. In certain situations, only the virtual-casing vector potential, or equivalently the virtual-casing poloidal magnetic flux function is required \citep{lao2005mhd,Marx_2017,pustovitov2021analytical}, in which case the method we presented in our first article is the most appropriate method. In other situations, knowledge of the component of the magnetic field normal to the plasma surface is desired \citep{landremanboozer,Landreman_REGCOIL}. In these cases, both our previous article and the present letter offer satisfactory solutions, although the approach presented here has the advantage of yielding the normal component of the magnetic field directly, bypassing the Fourier differentiation of the vector potential. Finally, there exist situations in which the tangential component of the magnetic field on the plasma surface is required, for example as a Neumann boundary condition for the poloidal magnetic flux function \citep{Zaitsev_2011,blum2012reconstruction,ricketson2016accurate}. The method we present here is uniquely well suited for these situations.

The letter is structured as follows. 
In section \ref{sec:vcp}, we review the virtual-casing principle for axisymmetric settings and highlight the relevant line integrals for computing the magnetic field due to the plasma, evaluated on the plasma boundary. Section \ref{sec:pv_high_order} describes numerical quadrature schemes for evaluating the principal value integrals that appear in these line integrals. In section \ref{sec:results}, we illustrate the performance of the quadrature schemes on an example determined by an analytic Grad-Shafranov equilibrium, and demonstrate the high order convergence and excellent accuracy of the Kapur-Rokhlin scheme. We conclude in section \ref{sec:conclusion}.

\section{Axisymmetric virtual-casing principle}\label{sec:vcp}

\subsection{Cylindrical coordinates for toroidally axisymmetric surfaces}

Throughout the letter, we will be concerned with plasmas with toroidally axisymmetric boundaries. We shall therefore make use of the standard, right-handed cylindrical coordinates $(r, \phi, z)$ naturally associated with the toroidal geometry, where the $z$-axis is the axis of revolution. At a point with toroidal angle $\phi$, we write the orthonormal unit vectors as $\vec e_r(\phi)$, $\vec e_\phi(\phi)$, and $\vec e_z$. With this notation, we emphasize the fact that the radial and azimuthal unit vectors depend on the toroidal angle. 

Let $\gamma$ be the simple closed curve in the $(r,z)$ plane such that by rotating $\gamma$ about the $z$-axis for $\phi \in [0, 2\pi]$, we obtain the closed surface of revolution $\Gamma$ corresponding to the plasma boundary. The generating curve $\gamma$ is parameterized by a single variable $t$, which we assume has period $L$. We denote the components of $\gamma$ in the $(r,z)$ plane by $(r(t), z(t))$, and we identify a point $\vec y \in \Gamma$ by its toroidal revolution angle $\phi$ and its generating curve parameter $t$. Correspondingly, we often write $\vec y = \vec y(\phi, t)$ to stress this parameterization. 
Moreover, we assume that $\gamma$ is a $C^1$ curve which does not intersect the $z$-axis, in the sense that the derivatives $r'(t)$ and $z'(t)$ are continuous on $[0,L]$ and there exists $R_{min}>0$ for which $r(t) \ge R_{min}$ on $[0,L]$.

\subsection{Virtual-casing principle for axisymmetric plasmas}

Consider an axisymmetric plasma confined by external coils in equilibrium. Let $\vec{B}$ denote the total magnetic field. The poloidal field $\vec B^{pol} = \vec{B} - B_{\phi} \vec{e}_{\phi}$ at any location is the sum of the poloidal field $\vec B_{ext}^{pol}$ due to the external coils, and of the poloidal field $\vec B_V^{pol}$ due to the plasma current. The field $\vec B_V^{pol}$ is given for all $\vec x \in \R^3$ by the Biot-Savart law:
\begin{equation}
    \vec B_V^{pol}(\vec x)=\frac{\mu_{0}}{4\pi} \iiint_{\Omega}J^{tor}_{V}(\vec y) 
    \vec{e}_\phi(\phi(\vec{y})) \cross \frac{\vec x - \vec y}{\norm{\vec x - \vec y}^3} \d \vec y
    \label{eq:volume_int}    
\end{equation}
where $\mu_{0}$ is the permeability of free space, the integration volume $\Omega$ is the plasma domain, and $J^{tor}_{V}$ is the toroidal current density in the plasma. The vector $\vec{e}_{\phi}(\phi(\vec{y}))$ is the unit vector in the toroidal direction at $\vec{y}$, where the argument $\phi(\vec{y})$ is the toroidal angle of the point $\vec{y}$.

Equation \eqref{eq:volume_int} is a volume integral, which is expensive to evaluate numerically. The virtual-casing principle gives a formula for $\vec B^{pol}_V$ that depends only on the full field $\vec B^{pol}$ at the plasma boundary, and only requires the evaluation of a surface integral (i.e. line integral for axisymmetric domains) \citep{Shafranov_1972,Zakharov_1973,hanson2015virtual}. 
Specifically, the virtual-casing principle states that if $\Gamma$ is the flux surface bounding the plasma, then $\vec B_V^{pol}$ can be written in terms of a field generated by the toroidal surface current $\vec J^{tor}_S$ such that $\mu_0 \vec J^{tor}_S = -\vec n \cross \vec B^{pol}$, where $\vec n$ is the outward unit normal vector to $\Gamma$. This relation is given according to \citet{hanson2015virtual}:
\begin{equation} \label{eq:virtual_casing}
    \vec B^{pol}_V(\vec x) = 
    \frac 1{4\pi} \iint_{\Gamma} \left[
    \frac{(\vec n(\vec y) \cross \vec B^{pol}(\vec y)) \cross (\vec x - \vec y)}{\norm{\vec x - \vec y}^3}
    \right] \d \Gamma(\vec y) 
    + 
    \begin{cases}
        \vec B^{pol}(\vec x) \quad &\vec x \in \Omega \\
        \vec B^{pol}(\vec x) / 2 \quad &\vec x \in \Gamma \\
        \vec 0 \quad &\vec x \notin \overline{\Omega}.
    \end{cases}
\end{equation}

In this letter, we are interested in the most common application of the virtual-casing principle, in which we want to compute the magnetic field due to the plasma currents \textit{on} the plasma boundary, i.e. for $\vec{x}\in\Gamma$. That motivates why we focus on the poloidal magnetic field. Indeed, for axisymmetric plasma boundaries and magnetic fields, the toroidal magnetic field on the plasma boundary is entirely generated by external coils. To see this, let $C_{r,z}$ be any circle of radius $r$ in a plane perpendicular to the $z$-axis, at altitude $z$, that is tangent to the axisymmetric plasma boundary. Applying Stokes' theorem along this circle, we can write
\begin{equation}
    \oint_{C_{r,z}}B_{\phi}(r,z)r \d\phi = 2\pi r B_{\phi}(r,z) = \mu_{0}\iint_{D_{r,z}} \vec{J} \vec{\cdot} \d\vec{S} = \mu_{0} I_{\mathrm{coils}}    
\end{equation}
where $D_{r,z}$ is the flat disc perpendicular to the $z$-axis bounded by the circle $C_{r,z}$, and $I_{\mathrm{coils}}$ is the total coil-induced current. In the last equality, we have used the fact that the plasma currents are divergence free. We obtain the well-known expression for the vacuum toroidal field, $B_{\phi}(r)=\mu_{0}I_{\mathrm{coils}}/(2\pi r)$, which holds all the way to the plasma boundary, and confirms that for axisymmetric plasmas, plasma currents do not contribute to the toroidal magnetic on the plasma boundary.

\subsection{Reduction of the virtual-casing principle to line integrals} \label{sec:reduce_1d}

Since we are considering an axisymmetric surface $\Gamma$, the surface integral in \eqref{eq:virtual_casing} can be expressed in terms of a line integral by integrating over the toroidal angle analytically.

The poloidal magnetic field ${\vec B}^{pol}$ at any point $\vec y(\phi, t) \in \Gamma$ can be expressed in terms of its poloidal flux function $\psi(r,z)$ and the parameterization $(\phi,t) \mapsto \vec y(\phi, t)$ by \citep{freidberg2014ideal}
\begin{align}
    \vec B^{pol}(\vec y(\phi,t)) 
    &= \grad \psi(r(t),z(t)) \cross \grad \phi \nonumber \\
    &= -\frac 1{r(t)} \pder\psi z(r(t), z(t)) \vec e_r(\phi) 
        + \frac 1{r(t)} \pder\psi r(r(t), z(t)) \vec e_z.
        \label{eq:B_axisym}
\end{align}

Inserting this expression in \eqref{eq:virtual_casing} and integrating it analytically with respect to the toroidal angle, we may evaluate $\vec{B}_V^{pol}$ directly through the remaining univariate integral. 
Presently, we consider evaluating on the plasma boundary $\Gamma$. Without loss of generality, we assume that the evaluation point lies on the cross-section of the plasma boundary with azimuthal angle $\phi=0$, which has cylindrical coordinate unit vectors that align with the rectangular unit vectors $\{\vec{e}_x, \vec{e}_y, \vec{e}_z\}$. We denote such an evaluation point as $(R,0,Z)$ in cylindrical coordinates.

Upon integrating the toroidal angle, the univariate expression for $\vec{B}_V^{pol}$ on the plasma boundary is:
\begin{equation} \label{eq:vcp_1d}
    \vec B_V^{pol}(R,Z)
    = \frac{1}{4\pi} \int_{0}^{L} \vec{f}(t) \d t
    + \frac{1}{2} \vec B^{pol}(R,Z),
    \qquad (R,0,Z) \in \Gamma,
\end{equation}
where \citep{toler_cerfon_malhotra_2023}
\begin{multline} \label{eq:vcp_integrand}
    \vec{f}(t)
    = \frac {2} {r(t) \sqrt{\alpha + \beta}}
    \left[ \pder\psi z r'(t) - \pder\psi r z'(t) \right]
    \Bigg\{ \frac {Z - z(t)}{R} 
        \left( 
        -K(k^2) 
        + \frac{\alpha}{\alpha - \beta} E(k^2) 
        \right) \vec{e}_x 
    \\
    + \left( 
        K(k^2) 
        + \frac{r(t)^2 - R^2 - (Z-z(t))^2}{\alpha - \beta} E(k^2) 
        \right) \vec{e}_z \Bigg\}.
\end{multline}
Here, the functions $K$ and $E$ are the complete elliptic integrals of the first and second kind, primes denote derivatives with respect to the parameterization variable $t$, and we have introduced the quantities
\begin{equation}
\begin{cases}
    \alpha  = \alpha(t; R,Z) = R^2 + r(t)^2 + (Z-z(t))^2 \\
    \beta   = \beta(t; R,Z)  = 2 R r(t) \\
    k^2     = k(t; R,Z)^2    = \displaystyle \frac{2\beta}{\alpha+\beta}.
\end{cases}
\end{equation}

Since we are evaluating on the plasma boundary, the evaluation point can be identified by its generating curve parameter $t$, say $t_0$, so $r(t_0)=R$ and $z(t_0)=Z$.
There are two sources of singularity in $\vec{f}(t)$ near $t_0$.
The mapping $t \mapsto K(k(t)^2)$ is $O(\log|t-t_0|)$ near $t_0$\citep{toler_cerfon_malhotra_2023,gradshteyn2014table}, and the errors from Taylor series for $r(t)$ and $z(t)$ yield $1/(\alpha-\beta) = O(1/(t-t_0)^2)$.

\section{High-order principal value quadrature for the axisymmetric virtual-casing principle} \label{sec:pv_high_order}

In \citet{toler_cerfon_malhotra_2023}, we showed that the integral in \eqref{eq:vcp_1d} is not convergent. That means that this integral must be interpreted in the Cauchy principal value sense for all components of $\vec B_V^{pol}(R,Z)$ in \eqref{eq:vcp_1d} to be well defined. Since
the observation point $(R,Z)$ lies on the plasma boundary, we may identify this point with a value of the boundary parameter $t$, say $t_0$.

In order to evaluate all the components of $\vec B_V^{pol}(R,Z)$ via \eqref{eq:vcp_1d}, we must evaluate the integral in this expression with the following substitution:
\begin{equation} \label{eq:pv_subst}
    \int_{0}^{L} \vec{f}(t) \d t
    \qquad \xrightarrow{\qquad} \qquad 
    \lim_{\epsilon \to 0^+} \left( 
        \int_{0}^{t_0-\epsilon} \vec{f}(t) \d t + \int_{t_0+\epsilon}^{L} \vec{f}(t) \d t \right),
\end{equation}
where $\vec{f}$ is given in \eqref{eq:vcp_integrand}. (If $t_0=0$, the integration interval within the limit should be $[\epsilon, L-\epsilon]$.)
Let us discuss two quadrature rules to do so: the alternating trapezoidal rule and the periodic Kapur-Rokhlin rule \citep{kapur1997high,toler_cerfon_malhotra_2023}.

We begin by considering the alternating trapezoidal rule. We recall that the standard $N$-point periodic trapezoidal rule for the integral $\int_{0}^{L} g(t) \d t$ for an $L$-periodic integrand $g$ is defined by equispaced quadrature nodes with uniform weights $h=L/N$.
The alternating trapezoidal rule for the case when $g$ has a singularity at $t_0$ is a specific case of the trapezoidal rule, where the quadrature nodes $\{s_i\}_{i=1}^{N}$ are placed to straddle $t_0$. That is, the nodes are $s_i = t_0 + (i - \frac{1}{2}) h$ for $i=1, \dots, N$.
The alternating trapezoidal rule for some types of Cauchy principal value integrals is supported by mathematical theory. 
In particular, it is known to converge spectrally accurately for principal value integrals of the form $\int_{-1}^1 (q(t)/t) \d t$ when $q$ is an analytic function \citep{trefethen2014exponentially,kress1970anwendung}. And if $q \in C^{P}[0,L]$ but is not analytic, then the quadrature converges with order $P$ \citep{Sidi1988}. 
By staggering quadrature nodes with equal weights on each side of the singularity, the alternating trapezoidal rule captures the cancellation that is necessary for a well-defined Cauchy principal value. For our application, we note that the asymptotic behavior of the integrand is not $q(t)/t$, but $q_1(t)/t + q_2(t) \log|t| + q_3(t)$ for bounded functions $q_1, q_2, q_3$.
As a result of the logarithmically singular term, we do not expect high-order convergence of the quadrature error. Rather, we expect low order convergence, as we will verify numerically in the next section.

In contrast, the Kapur-Rokhlin rule is precisely designed to integrate logarithmic singularities with high-order accuracy \citep{toler_cerfon_malhotra_2023}. We recall that the $(N-1)$-point rule for the periodic setting of interest in our context can be written as 
\begin{equation} \label{eq:KR_def}
   \int_{t_0-L/2}^{t_0+L/2} g(t) \d t 
    = \sum_{\substack{j=-(N-1)/2 \\ j \ne 0}}^{(N-1)/2} w_j g_j + O(h^n)
\end{equation}
where $g$ is logarithmically singular at $t_0$, with
\begin{equation} \label{eq:KR_weights}
    w_j = 
    \begin{cases}
        (1+\gamma_j + \gamma_{-j}) h, \quad & 1 \le |j| \le n \\
        h, \quad &\text{otherwise}
    \end{cases}
\end{equation}
and $g_j \equiv g(t_0 + j h)$.
The convergence order $n$ may be chosen by the user and the weights $\{ \gamma_j \}$ may be read from a table of values \citep{kapur1997high}.
The spacing $h=L/N$ between quadrature points is uniform (except for the doubled space between nodes $-1$ and $1$), and we are using notation so that the meaning of $h$ is the same for both the alternating trapezoidal rule and the Kapur-Rokhlin rule.
In \eqref{eq:KR_def}, we assume that $N$ is odd to simplify notation. If $N$ is even, the points may be indexed by $j=\pm 1, \dots, \pm(N/2-1), N/2$.

Like the alternating trapezoidal rule, the periodic Kapur-Rokhlin rule staggers nodes around the singularity with equal weights, so it also captures the principal value cancellation when integrating $\vec{f}$. Indeed, the weights in \eqref{eq:KR_weights} satisfy $w_{j} = w_{-j}$. However, the quadrature error converges much faster than for the alternating trapezoidal rule, since the quadrature rule was also designed for the logarithmic singularity of $\vec{f}$.
We next discuss practical performance of these numerical methods in the context of a virtual-casing principle calculation.

\section{Numerical results}\label{sec:results}

To compare and verify the numerical methods for calculating the principal value of the integral in \eqref{eq:vcp_1d}, we use the same reference data as \citet{toler_cerfon_malhotra_2023} on the same geometry. Specifically, the axisymmetric plasma boundary is given by the level set $\{ \psi=0 \}$ of the poloidal flux function given by
\begin{equation}
    \psi(r,z) = \frac{\kappa F_B}{2 R_0^3 q_0}
    \left[\frac 14 (r^2 - R_0^2)^2 + \frac 1{\kappa^2} r^2 z^2 - a^2 R_0^2 \right],
\end{equation}
which solves the Grad-Shafranov equation with the Solov'ev profiles $\mu_{0}p(\psi)=-[F_{B}(\kappa + 1/\kappa)/( R_0^{3}q_0)]\psi$ and $F(\psi)=F_B$, where $p(\psi)$ is the plasma pressure profile, and $F(\psi)=rB_{\phi}$, with $B_\phi$ the toroidal magnetic field \citep{lutjens1996chease,lee2015ecom}. The parameters $R_0$ and $q_0$ may be interpreted as the major radius and safety factor at the magnetic axis, and $\kappa$ and $a$ as the elongation and minor radius of the plasma boundary. Throughout, we use the fusion relevant values $F_B = R_0 = q_0 = 1$ and $\kappa = 1.7$ and $a = 1/3$. 
The level set $\{ \psi = 0 \}$ may be parameterized by the functions \citep{lutjens1996chease,lee2015ecom}
\begin{equation} \label{eq:lee_geom}
    (r(t))^2 = R_0^2 + 2 a R_0 \cos t 
    \quad \text{and} \quad 
    z(t) = \kappa a \frac{R_0}{r(t)} \sin t
\end{equation}
for $t \in [0, L]$ with $L=2\pi$.

\begin{figure}
    \centering
    \includegraphics[width=0.98\textwidth]{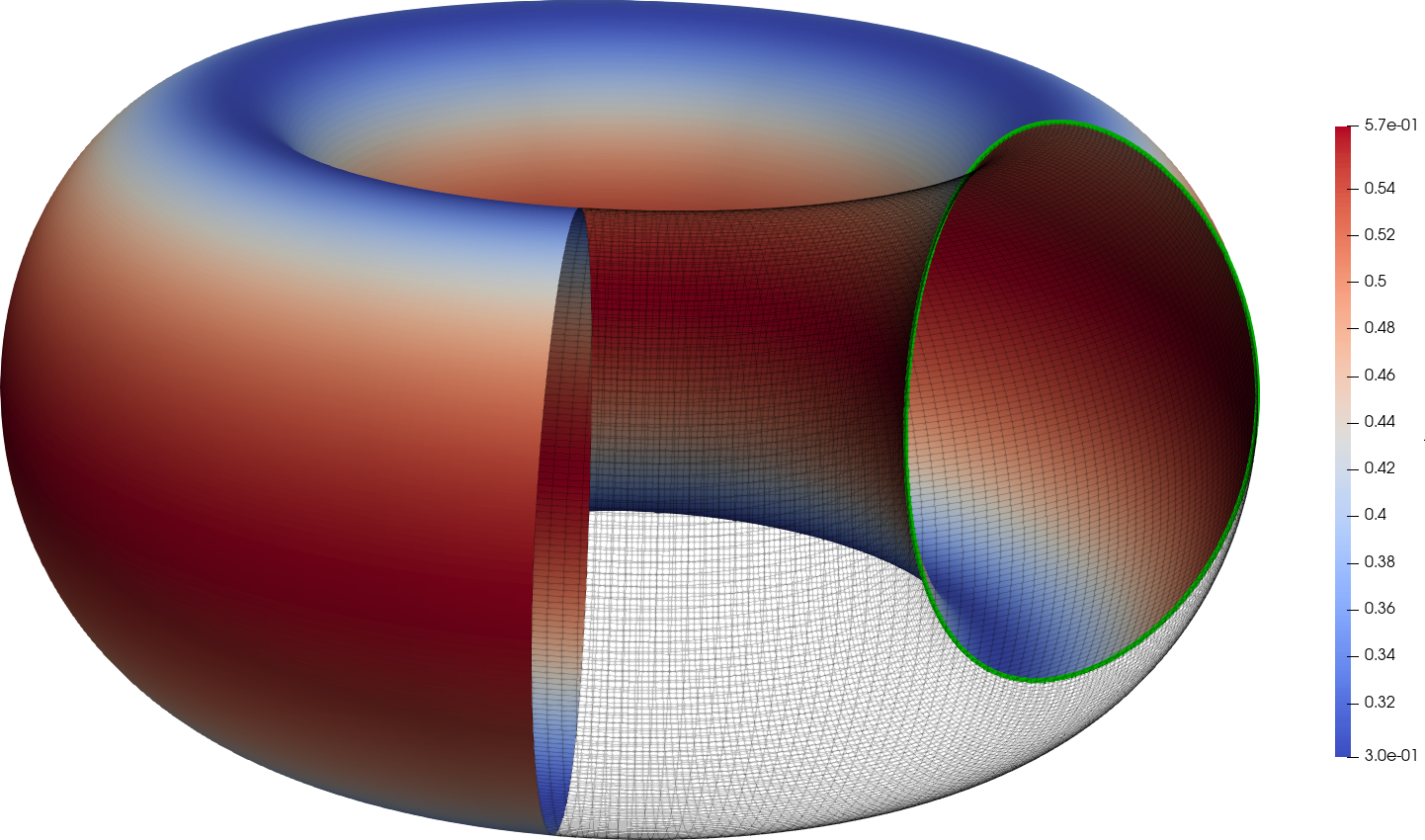}
    \caption{Reference solution for the Solov'ev profiles of section \ref{sec:results}. Colors represent $\norm{\vec{B}}$ on the plasma surface.}
    \label{fig:ref_sol}
\end{figure}

We take as ground truth the plasma-induced magnetic field from a high-resolution computation by the approach of \citet{malhotra2019efficient}.
The method used in that implementation is different from the one-dimensional integral approach presented here in several ways, making it appropriate for our verification.
Specifically, the code presented in \citet{malhotra2019efficient} views the plasma equilibrium as a fully three-dimensional equilibrium, and does not assume axisymmetry. Furthermore, \citet{malhotra2019efficient} obtain $\vec B^{pol}_V$ on the plasma boundary directly by evaluating the virtual-casing principle as a surface integral, as opposed to first reducing to a univariate integral by axisymmetry. Finally, the Cauchy principal value of the surface integral is numerically evaluated via a partition of unity scheme to handle the singularity of the integrand. 
The reference solution was computed on a fine grid with 1,580 toroidal angle values and 1,200 poloidal points on each cross-section. In a self-convergence comparison with a solution from finer resolution, this data was found to be accurate to 11 digits of precision. Figure \ref{fig:ref_sol} illustrates the geometry and reference solution for this problem.
This approach by \citet{malhotra2019efficient} is more robust by treating fully three-dimensional volumes with two-dimensional surfaces, but axisymmetric specializations like those we present here require discretization of one less dimension and result in faster runtimes.

With these reference values, we shall assess the numerical accuracy of the alternating trapezoidal rule and the Kapur-Rokhlin rule, which we discussed in the previous section.
For each quadrature scheme, we compute both the radial component $\vec B_{V,R}^{pol}$ of $\vec B_V^{pol}$ and the vertical component $\vec B_{V,Z}^{pol}$ at 1,200 points on the plasma boundary which are equispaced in the generating curve parameter $t$.
We measure the relative quadrature error for each component as the maximum of the pointwise absolute difference between the computed field component and the reference values.
We normalize both errors by the maximum absolute value of the 2,400 reference values for $\vec B_{V,R}^{pol}$ and $\vec B_{V,Z}^{pol}$.

\begin{figure}
    \centering
    \includegraphics[width=0.49\textwidth]{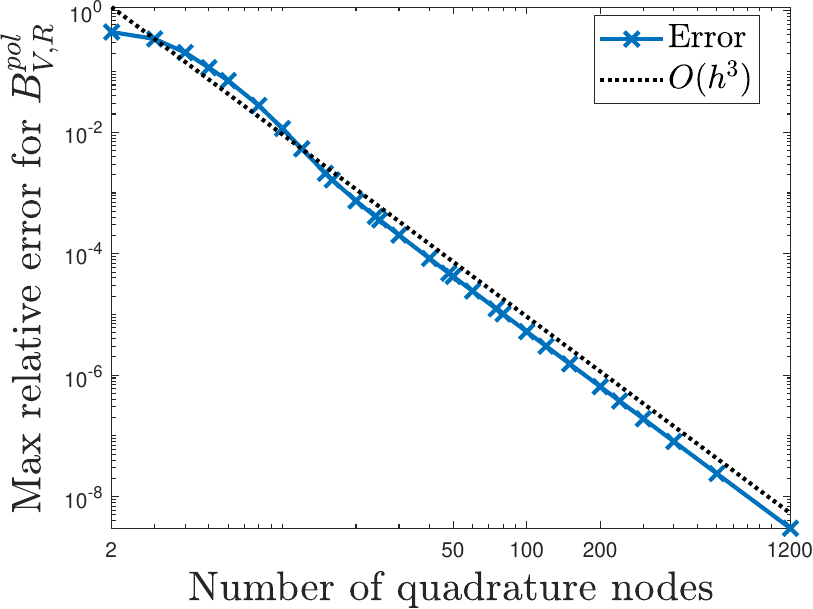}
    \includegraphics[width=0.49\textwidth]{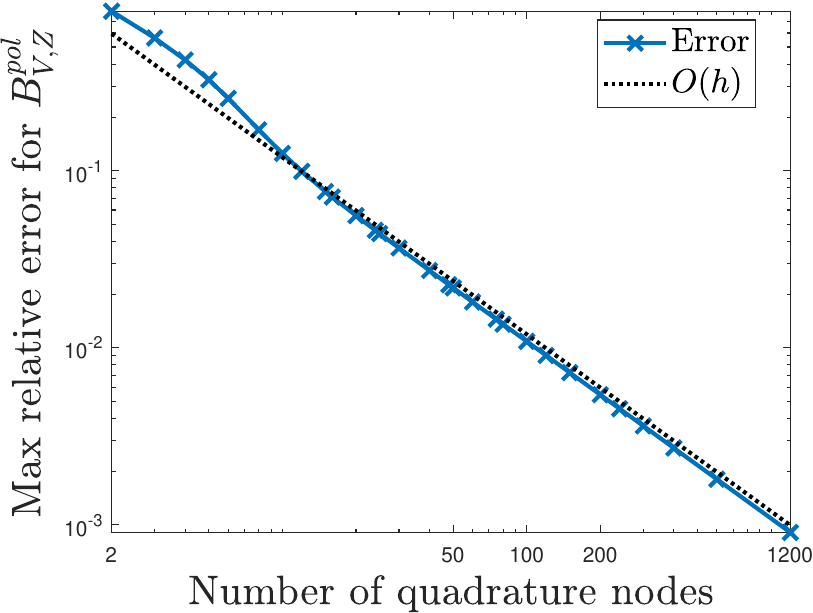}
    \caption{Maximum pointwise error using the alternating trapezoidal rule to compute $\vec B_{V,R}^{pol}$ and $\vec B_{V,Z}^{pol}$ via principal value.}
    \label{fig:pv_vcp_convergence_Jungpyo_AT}
\end{figure}

Figure \ref{fig:pv_vcp_convergence_Jungpyo_AT} illustrates that the alternating trapezoidal rule indeed converges to the same values found with the scheme presented in \citet{malhotra2019efficient}, but does so with a low order of convergence. 
The convergence rates are $O(h^3)$ for the radial component $\vec B_{V,R}^{pol}$ and $O(h)$ for the vertical component $\vec B_{V,Z}^{pol}$.
The radial component converges more quickly because its integrand has more regularity due to the following asymptotic behavior. Recall from section \ref{sec:reduce_1d} that $t \mapsto K(k(t)^2)$ is $O(\log|t-t_0|)$ near $t_0$. Hence in the integrand $\vec{f}$ of \eqref{eq:vcp_integrand}, the product $(Z-z(t))K(k^2)$ has only a removable discontinuity at $t_0$, and not a logarithmic singularity. The only singularity in the radial component is the principal value singularity of $(Z-z(t))/(\alpha-\beta)$. No such product dampens the singularity of $K$ in the vertical component. Nevertheless, the convergence rates for both components of $\vec B_{V}^{pol}$ are better than those guaranteed by the quadrature theory we discussed in section \ref{sec:pv_high_order}.

\begin{figure}
    \centering
    \includegraphics[width=0.49\textwidth]{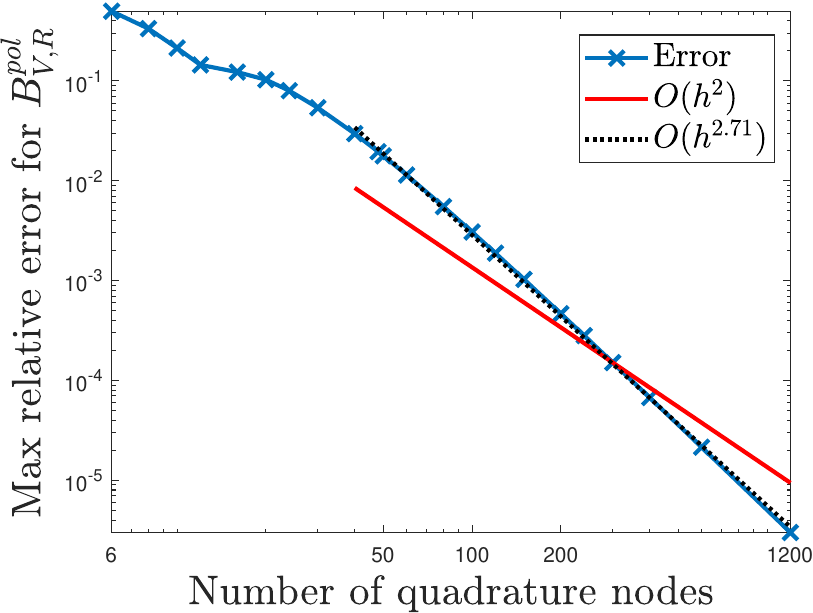}
    \includegraphics[width=0.49\textwidth]{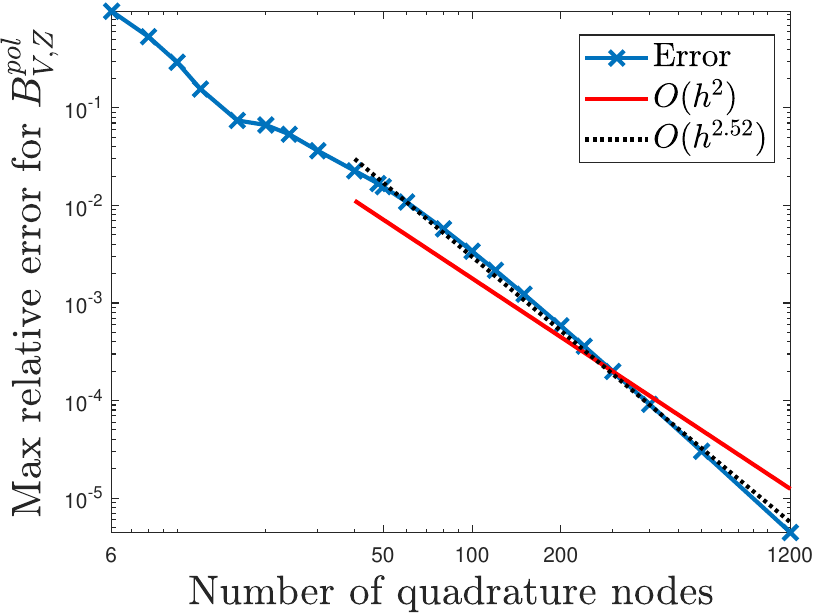}
    \caption{Maximum pointwise error using the 2nd order Kapur-Rokhlin rule to compute $\vec B_{V,R}^{pol}$ and $\vec B_{V,Z}^{pol}$ via principal value.}
    \label{fig:pv_vcp_convergence_Jungpyo_KR_order2}
\end{figure}

\begin{figure}
    \centering
    \includegraphics[width=0.49\textwidth]{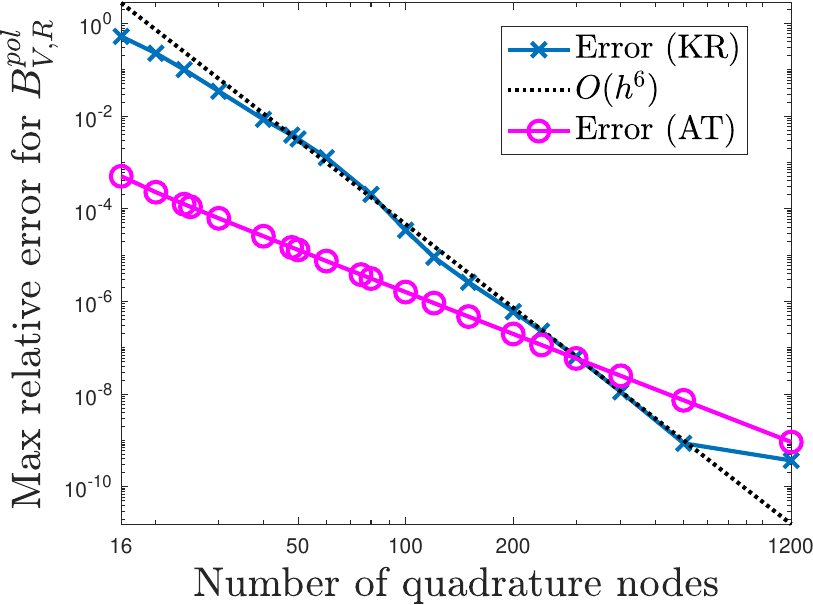}
    \includegraphics[width=0.49\textwidth]{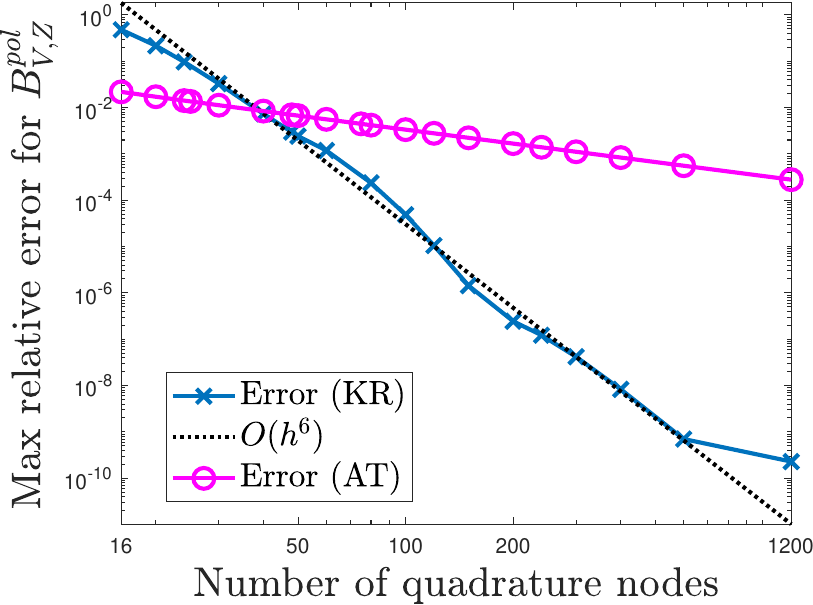}
    \caption{Maximum pointwise error using the 6th order Kapur-Rokhlin rule to compute $\vec B_{V,R}^{pol}$ and $\vec B_{V,Z}^{pol}$ via principal value. We compare the Kapur-Rokhlin (KR) error with the alternating trapezoidal (AT) error of figure \ref{fig:pv_vcp_convergence_Jungpyo_AT}.}
    \label{fig:pv_vcp_convergence_Jungpyo_KR_order6}
\end{figure}

\begin{figure}
    \centering
    \includegraphics[width=0.49\textwidth]{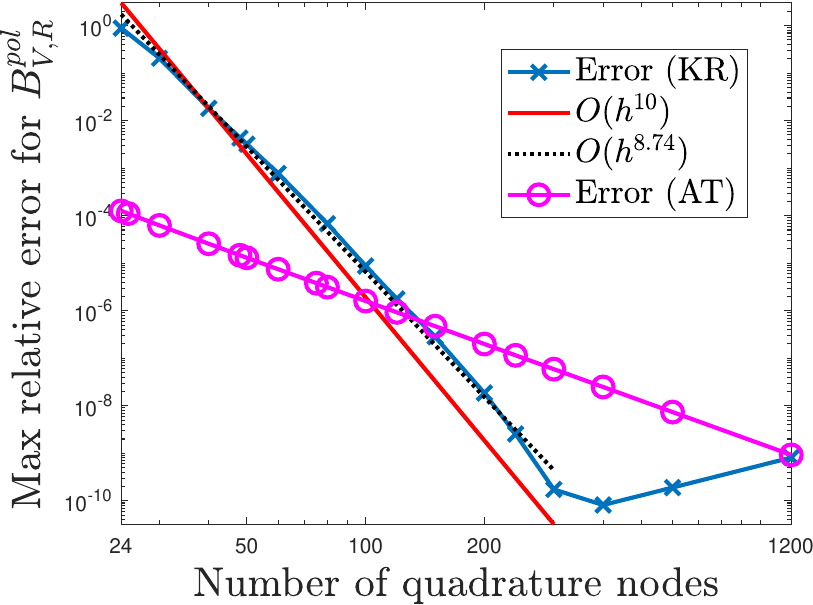}
    \includegraphics[width=0.49\textwidth]{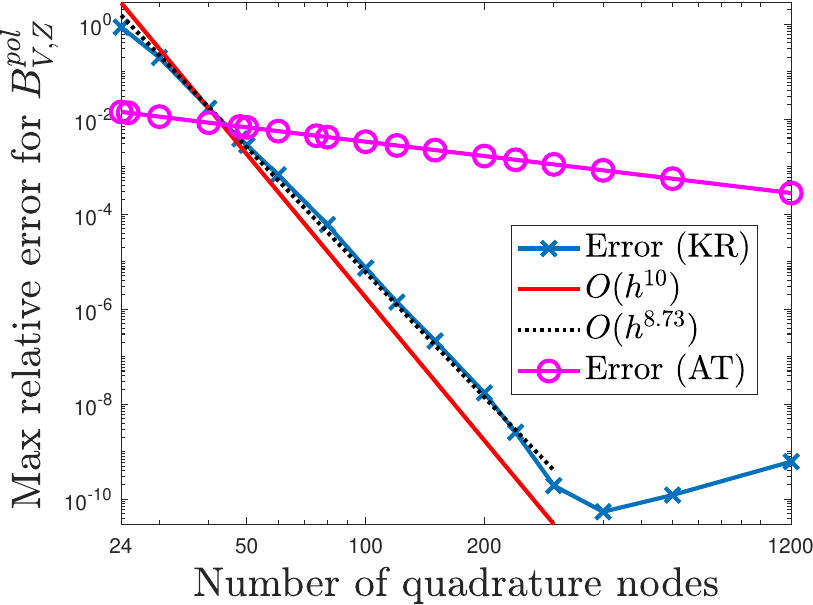}
    \caption{Maximum pointwise error using the 10th order Kapur-Rokhlin rule to compute $\vec B_{V,R}^{pol}$ and $\vec B_{V,Z}^{pol}$ via principal value. We compare the Kapur-Rokhlin (KR) error with the alternating trapezoidal (AT) error of figure \ref{fig:pv_vcp_convergence_Jungpyo_AT}.}
    \label{fig:pv_vcp_convergence_Jungpyo_KR_order10}
\end{figure}

We next consider the periodic Kapur-Rokhlin rule for the principal value quadrature.
We view each component of the integrand $\vec f$ as the sum of a logarithmically singular term and a remainder term:
\begin{equation} \label{eq:log_separate}
    \begin{cases}
        \vec e_x \boldsymbol{\cdot} \vec f(t) = p_R(t) \log|t-t_0| + q_R(t) \\
        \vec e_z \boldsymbol{\cdot} \vec f(t) = p_Z(t) \log|t-t_0| + q_Z(t).
    \end{cases}
\end{equation}
In particular, we have
\begin{align}
    \label{eq:pR}
    p_R(t) \log|t-t_0| &= \frac {2} {r(t) \sqrt{\alpha + \beta}}
        \left[ \pder\psi z r'(t) - \pder\psi r z'(t) \right]
        \frac {Z - z(t)}{R} \left( -K(k^2) \right) 
        \quad \text{and}
    \\
    \label{eq:pZ}
    p_Z(t) \log|t-t_0| &= \frac {2} {r(t) \sqrt{\alpha + \beta}}
        \left[ \pder\psi z r'(t) - \pder\psi r z'(t) \right]
        K(k^2) 
\end{align}
and we set $q_R$ and $q_Z$ to be the remaining corresponding terms in \eqref{eq:vcp_integrand}.
In figures \ref{fig:pv_vcp_convergence_Jungpyo_KR_order2}, \ref{fig:pv_vcp_convergence_Jungpyo_KR_order6}, and \ref{fig:pv_vcp_convergence_Jungpyo_KR_order10}, we show the results of applying the 2nd, 6th, and 10th order Kapur-Rokhlin rules, respectively. These are the theoretical convergence rates guaranteed for smooth coefficient functions $p_R, p_Z, q_R, q_Z$.
In this example, however, we see convergence rates that deviate slightly from these values.
The 6th order rule converges at the expected rate of $O(h^6)$. But, the 2nd order rule converges faster than predicted at rates of approximately $O(h^{2.71})$ and $O(h^{2.52})$ for the radial and vertical components, respectively. And the 10th order rule converges at approximate rates of $O(h^{8.74})$ and $O(h^{8.73})$.
We compute these approximate rates by performing an ordinary least squares line of best fit to the data measurements in the asymptotic regime of convergence.

Notably, the 10th order scheme converges slightly slower than predicted. We offer two potential explanations for this phenomenon. 
First, 
the smoothness assumptions of Kapur-Rokhlin are violated, as we explain presently. Similar to the alternating trapezoidal rule, the Kapur-Rokhlin rule requires that all of $\{p_R, p_Z, q_R, q_Z\}$ are sufficiently smooth to guarantee the theoretical convergence rate \citep{kapur1997high}. However, both $q_R$ and $q_Z$ include the elliptic integral $E$. The mapping $t \mapsto E(k(t)^2)$ is continuous, but its first derivative is not. This is readily seen by the identity of \citet{gradshteyn2014table} that
\begin{equation}
    \pder{E(k^2)}{k^2} = \frac{E(k^2) - K(k^2)}{2 k^2}.
\end{equation}
The first derivative of $t \mapsto E(k(t)^2)$, and hence the first derivatives of $q_R$ and $q_Z$, are logarithmically singular at $t_0$ due to the appearance of $K$.
Second,
very high orders of the Kapur-Rokhlin rule have weight corrections $\{\gamma_j\}$ which are large in magnitude and are sign-indefinite. This introduces numerical instability as a price to pay for the higher convergence order \citep{martinsson2019fast}. 

We hypothesize that the instability from the weight corrections also explains the observation that the quadrature error saturates at about nine digits of accuracy in figures \ref{fig:pv_vcp_convergence_Jungpyo_KR_order6} and \ref{fig:pv_vcp_convergence_Jungpyo_KR_order10}. It is numerically common for algorithmic instabilities to raise the error floor from machine precision. The instability is known to worsen as the theoretical convergence order increases, so we see this phenomenon earlier and more markedly for the 10th order rule compared to the 6th order rule. In contemporary plasma physics applications relevant to this work, however, accuracy at the level of eight or nine digits is typically not a limiting factor in overall calculations, so we do not explore the error saturation further.

Though the 2nd and 6th order rules have agreed with or exceeded the theoretical baselines in our experiment, we emphasize that we cannot guarantee such performance in general. The behavior of all of the Kapur-Rokhlin schemes in this setting are less predictable, but nevertheless promising in practice.

When we directly compare the performance of the Kapur-Rokhlin scheme and the alternating trapezoidal scheme in figures \ref{fig:pv_vcp_convergence_Jungpyo_KR_order6} and \ref{fig:pv_vcp_convergence_Jungpyo_KR_order10}, we see that both methods are competitive for computing $\vec{B}^{pol}_{V,R}$, whose virtual-casing integrand contains a less severe singularity. The Kapur-Rokhlin rule is asymptotically faster to converge, but the alternating trapezoidal rule has better accuracy when the number of quadrature points is small. However, the Kapur-Rokhlin rule more clearly outperforms the alternating trapezoidal rule for computing $\vec{B}^{pol}_{V,Z}$, whose virtual-casing integrand contains the more severe singularity. In this case, the alternating trapezoidal rule only has higher accuracy when the number of quadrature points is very limited: less than about 50.

\begin{figure}
    \centering
    \includegraphics[width=0.49\textwidth]{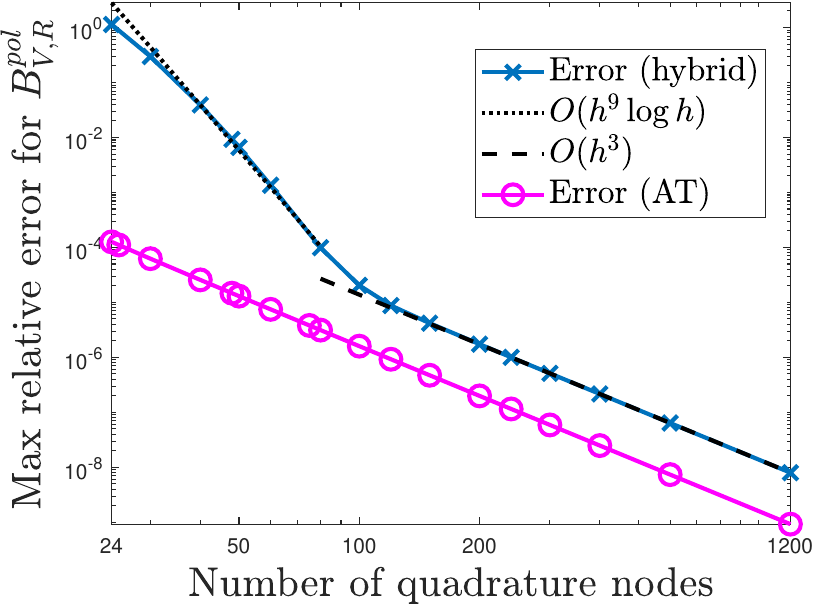}
    \includegraphics[width=0.49\textwidth]{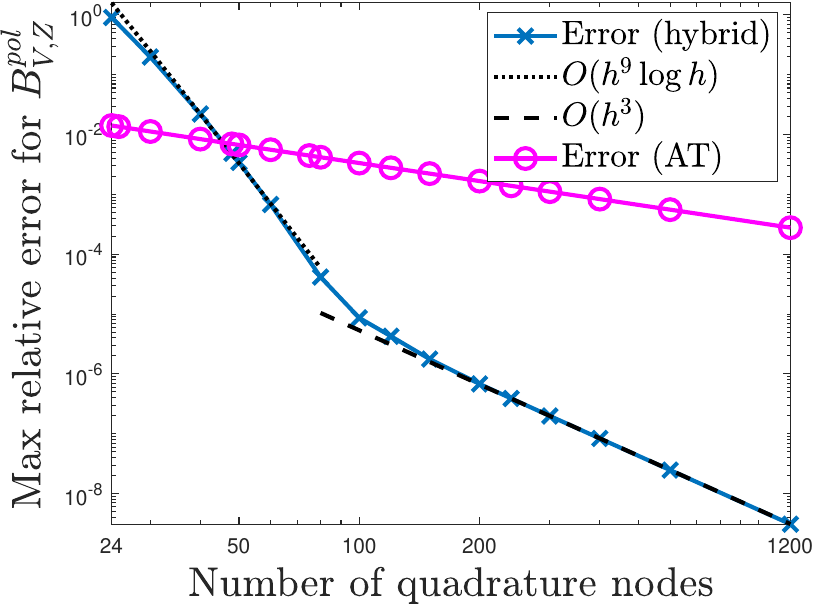}
    \caption{Maximum pointwise error using a combination of the 10th order Kapur-Rokhlin rule and the alternating trapezoidal rule to compute $\vec B_{V,R}^{pol}$ and $\vec B_{V,Z}^{pol}$. We compare the error with the alternating trapezoidal (AT) error of figure \ref{fig:pv_vcp_convergence_Jungpyo_AT}.}
    \label{fig:pv_vcp_convergence_Jungpyo_KR_order10_AT}
\end{figure}

Finally, we consider a third quadrature scheme which directly combines the alternating trapezoidal rule and the Kapur-Rokhlin rule. Returning to the decomposition \eqref{eq:log_separate}, we integrate the logarithmically singular terms with $p_R$ and $p_Z$ of \eqref{eq:pR} and \eqref{eq:pZ} via the periodic Kapur-Rokhlin rule, and we integrate the remaining terms $q_R$ and $q_Z$ via the alternating trapezoidal rule. 
One might imagine that this scheme would outperform both the pure Kapur-Rokhlin rule and the pure alternating trapezoidal rule, since this combined scheme uses each constituent quadrature on the term for which it is specialized. However, figure \ref{fig:pv_vcp_convergence_Jungpyo_KR_order10_AT} reveals that the lack of smoothness in the integrand again stalls convergence. 
For few quadrature nodes, the Kapur-Rokhlin corrections considerably improve convergence, but for more than about 100 nodes, the slower convergence of the alternating trapezoidal rule dominates. We observe that the limiting asymptotic convergence rate appears to be third-order convergence in this experiment for both the radial and vertical components of $\vec B_{V}^{pol}$. 
This improves the asymptotic rate from the pure alternating trapezoidal rule for $\vec B_{V,Z}^{pol}$, indicating that the alternating trapezoidal rule originally achieved only $O(h)$ convergence because it ignored the logarithmic singularity.

We additionally note that the asymptotic rates for the 2nd and 10th order Kapur-Rokhlin rules in figures 2 and 4 experimentally appear to agree with rates of $O(h^3 \log h)$ and $O(h^9 \log h)$ respectively, though this is our hypothesis and we are not aware of quadrature theory which would imply the presence of a $\log h$ factor in the convergence rate in this context. However, convergence orders with $\log h$ factors are known to appear when ill-suited quadrature rules, like the alternating trapezoidal rule, are used to integrate a function with a purely logarithmic singularity \citep{martinsson2019fast}. We display this feature for the 10th order rule in figure 5.

The pure Kapur-Rokhlin method remains the most efficient quadrature scheme discussed for this principle value application. It empirically displays high-order convergence and accuracy close to the theorized rates. The 10th order rule achieves a maximal accuracy of about nine digits with about 400 quadrature nodes.

\section{Conclusion}\label{sec:conclusion}
We have shown that both the alternating trapezoidal rule and the Kapur-Rokhlin rule are convergent schemes for the computation of {\em all} the components of $\vec B_{V}^{pol}$ expressed as one-dimensional Cauchy principal value integrals of the virtual-casing principle for axisymmetric equilibria. However, the convergence of the alternating trapezoidal rule is limited to very low order, which can be as low as one, because of the lack of smoothness of the integrand. In contrast, high order Kapur-Rokhlin schemes, which are nearly as easy to implement as the alternating trapezoidal rule, provide high order convergence for this calculation because the schemes are specifically designed for the singularities encountered in this application. 

We therefore recommend the direct implementation of high-order Kapur-Rokhlin rules for the computation of the axisymmetric virtual-casing principle as presented in this letter, since it is easier to use than the method originally proposed in \citep{toler_cerfon_malhotra_2023}, it provides all the components of the magnetic field, and the loss of accuracy is negligible as compared to the results obtained in \citep{toler_cerfon_malhotra_2023} for the normal component of the magnetic field only. 

\section*{Acknowledgements}
The authors would like to thank Mike O'Neil for insightful discussions. Evan Toler was supported by the National Science Foundation Graduate Research Fellowship under grant no. 1839302.
This work was supported by the U.S.~Department of Energy, Office of Science,
Office of Advanced Scientific Computing Research, Scientific Discovery through
Advanced Computing (SciDAC) Program through the FASTMath Institute under
Contract No.~DE-AC02-06CH11357.

\section*{Declaration of Interests}
The authors report no conflict of interest.

\bibliographystyle{plainnat}
\bibliography{biblio}

\begin{thebibliography}{26}
\providecommand{\natexlab}[1]{#1}
\providecommand{\url}[1]{\texttt{#1}}
\expandafter\ifx\csname urlstyle\endcsname\relax
  \providecommand{\doi}[1]{doi: #1}\else
  \providecommand{\doi}{doi: \begingroup \urlstyle{rm}\Url}\fi

\bibitem[Blum et~al.(2012)Blum, Boulbe, and Faugeras]{blum2012reconstruction}
Jacques Blum, Cedric Boulbe, and Blaise Faugeras.
\newblock Reconstruction of the equilibrium of the plasma in a tokamak and
  identification of the current density profile in real time.
\newblock \emph{Journal of computational physics}, 231\penalty0 (3):\penalty0
  960--980, 2012.

\bibitem[Drevlak et~al.(2018)Drevlak, Beidler, Geiger, Helander, and
  Turkin]{Drevlak_2018}
M.~Drevlak, C.D. Beidler, J.~Geiger, P.~Helander, and Y.~Turkin.
\newblock Optimisation of stellarator equilibria with {ROSE}.
\newblock \emph{Nuclear Fusion}, 59\penalty0 (1):\penalty0 016010, nov 2018.
\newblock \doi{10.1088/1741-4326/aaed50}.
\newblock URL \url{https://doi.org/10.1088/1741-4326/aaed50}.

\bibitem[Freidberg(2014)]{freidberg2014ideal}
Jeffrey~P Freidberg.
\newblock \emph{Ideal MHD}.
\newblock Cambridge University Press, 2014.

\bibitem[Gradshteyn and Ryzhik(2014)]{gradshteyn2014table}
Izrail~Solomonovich Gradshteyn and Iosif~Moiseevich Ryzhik.
\newblock \emph{Table of integrals, series, and products}.
\newblock Academic press, 2014.

\bibitem[Hanson(2015)]{hanson2015virtual}
James~D Hanson.
\newblock The virtual-casing principle and {Helmholtz’s} theorem.
\newblock \emph{Plasma Physics and Controlled Fusion}, 57\penalty0
  (11):\penalty0 115006, 2015.

\bibitem[Kappel et~al.(2023)Kappel, Landreman, and
  Malhotra]{kappel2023magnetic}
John Kappel, Matt Landreman, and Dhairya Malhotra.
\newblock The magnetic gradient scale length explains why certain plasmas
  require close external magnetic coils, 2023.

\bibitem[Kapur and Rokhlin(1997)]{kapur1997high}
Sharad Kapur and Vladimir Rokhlin.
\newblock High-order corrected trapezoidal quadrature rules for singular
  functions.
\newblock \emph{SIAM Journal on Numerical Analysis}, 34\penalty0 (4):\penalty0
  1331--1356, 1997.

\bibitem[Kress and Martensen(1970)]{kress1970anwendung}
R~Kress and E~Martensen.
\newblock Anwendung der rechteckregel auf die reelle hilberttransformation mit
  unendlichem intervall.
\newblock \emph{ZAMM-Journal of Applied Mathematics and Mechanics/Zeitschrift
  f{\"u}r Angewandte Mathematik und Mechanik}, 50\penalty0 (1-4):\penalty0
  61--64, 1970.

\bibitem[Landreman(2017)]{Landreman_REGCOIL}
Matt Landreman.
\newblock An improved current potential method for fast computation of
  stellarator coil shapes.
\newblock \emph{Nuclear Fusion}, 57\penalty0 (4):\penalty0 046003, feb 2017.
\newblock \doi{10.1088/1741-4326/aa57d4}.
\newblock URL \url{https://doi.org/10.1088/1741-4326/aa57d4}.

\bibitem[Landreman and Boozer(2016)]{landremanboozer}
Matt Landreman and Allen~H. Boozer.
\newblock Efficient magnetic fields for supporting toroidal plasmas.
\newblock \emph{Physics of Plasmas}, 23\penalty0 (3):\penalty0 032506, 2016.
\newblock \doi{10.1063/1.4943201}.
\newblock URL \url{https://doi.org/10.1063/1.4943201}.

\bibitem[Lao et~al.(2005)Lao, John, Peng, Ferron, Strait, Taylor, Meyer, Zhang,
  and You]{lao2005mhd}
LL~Lao, HE~St John, Q~Peng, JR~Ferron, EJ~Strait, TS~Taylor, WH~Meyer, C~Zhang,
  and KI~You.
\newblock Mhd equilibrium reconstruction in the diii-d tokamak.
\newblock \emph{Fusion science and technology}, 48\penalty0 (2):\penalty0
  968--977, 2005.

\bibitem[Lazerson et~al.(2013)Lazerson, Sakakibara, and Suzuki]{Lazerson_2013}
S~A Lazerson, S~Sakakibara, and Y~Suzuki.
\newblock A magnetic diagnostic code for 3d fusion equilibria.
\newblock \emph{Plasma Physics and Controlled Fusion}, 55\penalty0
  (2):\penalty0 025014, jan 2013.
\newblock \doi{10.1088/0741-3335/55/2/025014}.
\newblock URL \url{https://doi.org/10.1088/0741-3335/55/2/025014}.

\bibitem[Lee and Cerfon(2015)]{lee2015ecom}
Jungpyo Lee and Antoine Cerfon.
\newblock {ECOM}: A fast and accurate solver for toroidal axisymmetric {MHD}
  equilibria.
\newblock \emph{Computer Physics Communications}, 190:\penalty0 72--88, 2015.

\bibitem[L{\"u}tjens et~al.(1996)L{\"u}tjens, Bondeson, and
  Sauter]{lutjens1996chease}
Hinrich L{\"u}tjens, Anders Bondeson, and Olivier Sauter.
\newblock The chease code for toroidal mhd equilibria.
\newblock \emph{Computer physics communications}, 97\penalty0 (3):\penalty0
  219--260, 1996.

\bibitem[Malhotra et~al.(2019)Malhotra, Cerfon, O’Neil, and
  Toler]{malhotra2019efficient}
Dhairya Malhotra, Antoine~J Cerfon, Michael O’Neil, and Evan Toler.
\newblock Efficient high-order singular quadrature schemes in magnetic fusion.
\newblock \emph{Plasma Physics and Controlled Fusion}, 62\penalty0
  (2):\penalty0 024004, 2019.

\bibitem[Martinsson(2019)]{martinsson2019fast}
Per-Gunnar Martinsson.
\newblock \emph{Fast Direct Solvers for Elliptic PDEs}.
\newblock Society for Industrial and Applied Mathematics, Philadelphia, PA,
  2019.
\newblock \doi{10.1137/1.9781611976045}.
\newblock URL \url{https://epubs.siam.org/doi/abs/10.1137/1.9781611976045}.

\bibitem[Marx and Lütjens(2017)]{Marx_2017}
A~Marx and H~Lütjens.
\newblock Free-boundary simulations with the {XTOR}-2f code.
\newblock \emph{Plasma Physics and Controlled Fusion}, 59\penalty0
  (6):\penalty0 064009, may 2017.
\newblock \doi{10.1088/1361-6587/aa6d1f}.
\newblock URL \url{https://doi.org/10.1088/1361-6587/aa6d1f}.

\bibitem[Pustovitov and Chukashev(2021)]{pustovitov2021analytical}
VD~Pustovitov and NV~Chukashev.
\newblock Analytical solution to external equilibrium problem for plasma with
  elliptic cross section in a tokamak.
\newblock \emph{Plasma Physics Reports}, 47\penalty0 (10):\penalty0 956--966,
  2021.

\bibitem[Ricketson et~al.(2016)Ricketson, Cerfon, Rachh, and
  Freidberg]{ricketson2016accurate}
Lee~F Ricketson, Antoine~J Cerfon, Manas Rachh, and Jeffrey~P Freidberg.
\newblock Accurate derivative evaluation for any grad--shafranov solver.
\newblock \emph{Journal of Computational Physics}, 305:\penalty0 744--757,
  2016.

\bibitem[Shafranov and Zakharov(1972)]{Shafranov_1972}
V.D. Shafranov and L.E. Zakharov.
\newblock Use of the virtual-casing principle in calculating the containing
  magnetic field in toroidal plasma systems.
\newblock \emph{Nuclear Fusion}, 12\penalty0 (5):\penalty0 599--601, sep 1972.
\newblock \doi{10.1088/0029-5515/12/5/009}.
\newblock URL \url{https://doi.org/10.1088/0029-5515/12/5/009}.

\bibitem[Sidi and Israeli(1988)]{Sidi1988}
Avram Sidi and Moshe Israeli.
\newblock Quadrature methods for periodic singular and weakly singular
  {Fredholm} integral equations.
\newblock \emph{Journal of Scientific Computing}, 3\penalty0 (2):\penalty0
  201–231, June 1988.
\newblock ISSN 1573-7691.
\newblock \doi{10.1007/bf01061258}.
\newblock URL \url{http://dx.doi.org/10.1007/BF01061258}.

\bibitem[Toler et~al.(2023)Toler, Cerfon, and
  Malhotra]{toler_cerfon_malhotra_2023}
Evan Toler, A.J. Cerfon, and D.~Malhotra.
\newblock A fast, accurate and easy to implement {Kapur–Rokhlin} quadrature
  scheme for singular integrals in axisymmetric geometries.
\newblock \emph{Journal of Plasma Physics}, 89\penalty0 (2):\penalty0
  905890210, 2023.
\newblock \doi{10.1017/S002237782300020X}.

\bibitem[Trefethen and Weideman(2014)]{trefethen2014exponentially}
Lloyd~N Trefethen and JAC Weideman.
\newblock The exponentially convergent trapezoidal rule.
\newblock \emph{SIAM Review}, 56\penalty0 (3):\penalty0 385--458, 2014.

\bibitem[Zaitsev et~al.(2011)Zaitsev, Kostomarov, Suchkov, Drozdov, Solano,
  Murari, Matejcik, Hawkes, and Contributors]{Zaitsev_2011}
F.S. Zaitsev, D.P. Kostomarov, E.P. Suchkov, V.V. Drozdov, E.R. Solano,
  A.~Murari, S.~Matejcik, N.C. Hawkes, and JET-EFDA Contributors.
\newblock Analyses of substantially different plasma current densities and
  safety factors reconstructed from magnetic diagnostics data.
\newblock \emph{Nuclear Fusion}, 51\penalty0 (10):\penalty0 103044, sep 2011.
\newblock \doi{10.1088/0029-5515/51/10/103044}.
\newblock URL \url{https://dx.doi.org/10.1088/0029-5515/51/10/103044}.

\bibitem[Zakharov(1973)]{Zakharov_1973}
L.E. Zakharov.
\newblock Numerical methods for solving some problems of the theory of plasma
  equilibrium in toroidal configurations.
\newblock \emph{Nuclear Fusion}, 13\penalty0 (4):\penalty0 595--602, aug 1973.
\newblock \doi{10.1088/0029-5515/13/4/012}.
\newblock URL \url{https://doi.org/10.1088/0029-5515/13/4/012}.

\bibitem[Zakharov and Pletzer(1999)]{ZakharovPletzer}
L.E. Zakharov and A.~Pletzer.
\newblock Theory of perturbed equilibria for solving the grad–shafranov
  equation.
\newblock \emph{Physics of Plasmas}, 6\penalty0 (12):\penalty0 4693--4704,
  1999.
\newblock \doi{10.1063/1.873756}.
\newblock URL \url{https://doi.org/10.1063/1.873756}.

\end{thebibliography}

\end{document}